\documentclass[sigconf,nonacm]{acmart}

\usepackage{amsmath}
\usepackage{booktabs}
\usepackage{colortbl}
\usepackage{enumitem}
\usepackage{placeins}
\usepackage{graphicx}
\usepackage{multirow}
\usepackage{subcaption}
\usepackage[nameinlink,noabbrev]{cleveref}

\crefname{equation}{Eq.}{Eqs.}
\Crefname{equation}{Equation}{Equations}

\newcommand{\ie}{\emph{i.e., }}
\newcommand{\eg}{\emph{e.g., }}

\newcommand{\cf}{\emph{c.f. }}

\graphicspath{{figures/}}

\setcopyright{none}
\renewcommand\footnotetextcopyrightpermission[1]{}

\acmConference[Conference acronym 'XX]
  {Conference acronym 'XX}
  {June 03--05, 2018}
  {Woodstock, NY}

\AtBeginDocument{%
  }
\newcommand{\imprcell}[1]{\textcolor[HTML]{C00000}{\textbf{#1}}}

\begin{document}

\title{IMFuse: Instance-Aware Multi-Layer Fusion for LLM-Enhanced Sequential Recommendation}

\author{Yuheng Zheng}
\authornote{Both authors contributed equally to this research.}
\affiliation{%
  \institution{Zhejiang University}
  \city{Hangzhou}
  \country{China}}
\email{zhengyuheng@zju.edu.cn}

\author{Yu Cui}
\authornotemark[1]
\orcid{0009-0001-6203-3022}
\affiliation{%
  \institution{Zhejiang University}
  \city{Hangzhou}
  \country{China}}
\email{cuiyu23@zju.edu.cn}

\author{Bin Wu}
\affiliation{%
  \institution{Zhengzhou University}
  \city{Zhengzhou}
  \country{China}}
\email{wubin7019088@gmail.com}

\author{Jian Zhang}
\affiliation{%
  \institution{University of Science and Technology of China}
  \city{Hefei}
  \country{China}}
\email{zhangjian833@mail.ustc.edu.cn}

\author{Ye Feng}
\affiliation{%
  \institution{University of Science and Technology of China}
  \city{Hefei}
  \country{China}}
\email{yefengustc@mail.ustc.edu.cn}

\author{Can Wang}
\orcid{0000-0002-5890-4307}
\affiliation{%
  \institution{Zhejiang University}
  \city{Hangzhou}
  \country{China}}
\email{wcan@zju.edu.cn}

\author{Jiawei Chen}
\authornote{Corresponding author.}
\orcid{0000-0002-4752-2629}
\affiliation{%
  \institution{Zhejiang University}
  \city{Hangzhou}
  \country{China}}
\email{sleepyhunt@zju.edu.cn}

\renewcommand{\shortauthors}{Zheng et al.}

\begin{abstract}

Leveraging large language models (LLMs) to enrich item embeddings with textual semantics has become a prevailing paradigm for enhancing sequential recommendation (SR). However, existing methods typically rely solely on the final-layer LLM embeddings, leaving the semantics encoded in other layers largely unexploited.   Through an empirical analysis, we uncover three key findings: 1) item embeddings from deeper LLM layers suffer from increasingly severe spectral collapse, which reduces their information richness and hinders the optimization; 2) intermediate LLM layers encode complementary item semantics beyond the final layer; 3) different items exhibit heterogeneous layer-wise semantic patterns. These findings suggest that the final layer alone is insufficient and that multi-layer LLM semantics should be exploited adaptively for each item. 

To this end, we propose IMFuse, a novel Instance-Aware Multi-Layer Fusion strategy for LLM-enhanced SR. IMFuse adaptively aggregates embeddings from all LLM layers with learnable layer-wise fusion weights, and further customizes the fusion for each item via an instance-aware Mixture-of-Experts (MoE) module. As a plug-and-play strategy, IMFuse can be seamlessly integrated into various SR backbones and LLM-based enhancement methods. Extensive experiments on four real-world datasets demonstrate that IMFuse consistently improves state-of-the-art methods, achieving an average relative improvement of 6.72\% with limited computational overhead.

\end{abstract}



\maketitle

\section{Introduction}
\label{sec:introduction}


Sequential recommendation (SR), which aims to predict the next item a user will interact with based on the user's historical interaction sequence, plays a crucial role in many personalized services~\cite{hidasi2016gru4rec,kang2018self,sun2019bert4rec}. The key to SR lies in learning informative item embeddings that capture item characteristics and support accurate inference of user preferences. Conventional SR models are predominantly ID-based, assigning each item a unique ID and learning its embedding from user–item interactions. While effective at capturing collaborative signals, such ID-based models rely solely on interaction data and thus largely overlook the rich semantic information conveyed by item textual descriptions (\eg titles and categories), which can provide valuable complementary knowledge for recommendation.

To compensate for the semantic deficiency of ID embeddings, a growing body of research has turned to large language models (LLMs), whose pretraining on massive corpora equips them with rich world knowledge and strong text comprehension abilities \cite{grattafiori2024llama,brown2020language}. The prevailing paradigm encodes item texts into semantic embeddings via an LLM and then integrates them with ID embeddings in SR models. Despite their effectiveness, existing methods typically rely solely on the final-layer embeddings of LLMs, which may limit the full potential of LLM-based enhancement. We argue that this limitation is particularly pronounced in recommendation, as supported by the following empirical observations:
\begin{figure}[t]
  \centering

  \includegraphics[width=\columnwidth]{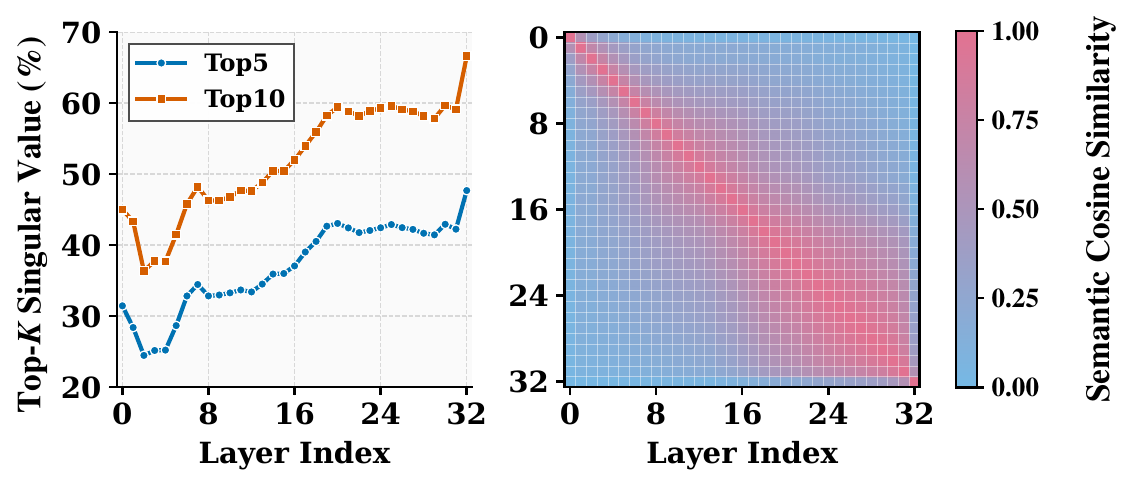}


  \caption{
  The Top-$K$ singular value proportion (left) and inter-layer cosine similarity analyses (right) with LLaMA-3-8B~\cite{grattafiori2024llama} as the LLM backbone.  Deeper LLM layer embeddings exhibit stronger spectral collapse, and different LLM layers exhibit significant semantic disparities.
  }

  \label{fig:layer-motivation}
  \vspace{-0.3cm}
\end{figure}



\textbf{1) Item embeddings from deeper LLM layers suffer from increasingly severe spectral collapse.}
 We quantify the spectral collapse ~\cite{hu2025alphafuse,cui2026spectran} of item representations at each layer by the proportion of the top-K singular values in the overall spectrum (\cf Figure~\ref{fig:layer-motivation} left). As the layer depth increases, spectral collapse becomes increasingly severe. Such collapse is harmful in several respects: the semantic signals of the embeddings become concentrated in a few dominant directions, reducing their information richness. Moreover, the collapsed representations may hinder model optimization and impair personalized preference modeling~\cite{lin2025popularity,yang2026pitfall}.

\textbf{2) Different LLM layers encode complementary item semantic information.}
Intuitively, embeddings from different LLM layers result from different depths of processing of the raw item semantics, and thus naturally characterize items at varying semantic granularities \cite{jawahar2019bert,tenney2019bert,liu2019linguistic}. Beyond the final layer, intermediate-layer embeddings may therefore carry complementary semantics that also benefit recommendation. Figure~\ref{fig:layer-motivation} (right) further visualizes the similarity between representations from different layers: while adjacent layers remain highly correlated, distant layers exhibit substantial differences, indicating that they capture diverse semantic granularities rather than redundant features.

\begin{figure}[t]
    \centering
    \includegraphics[width=\linewidth]{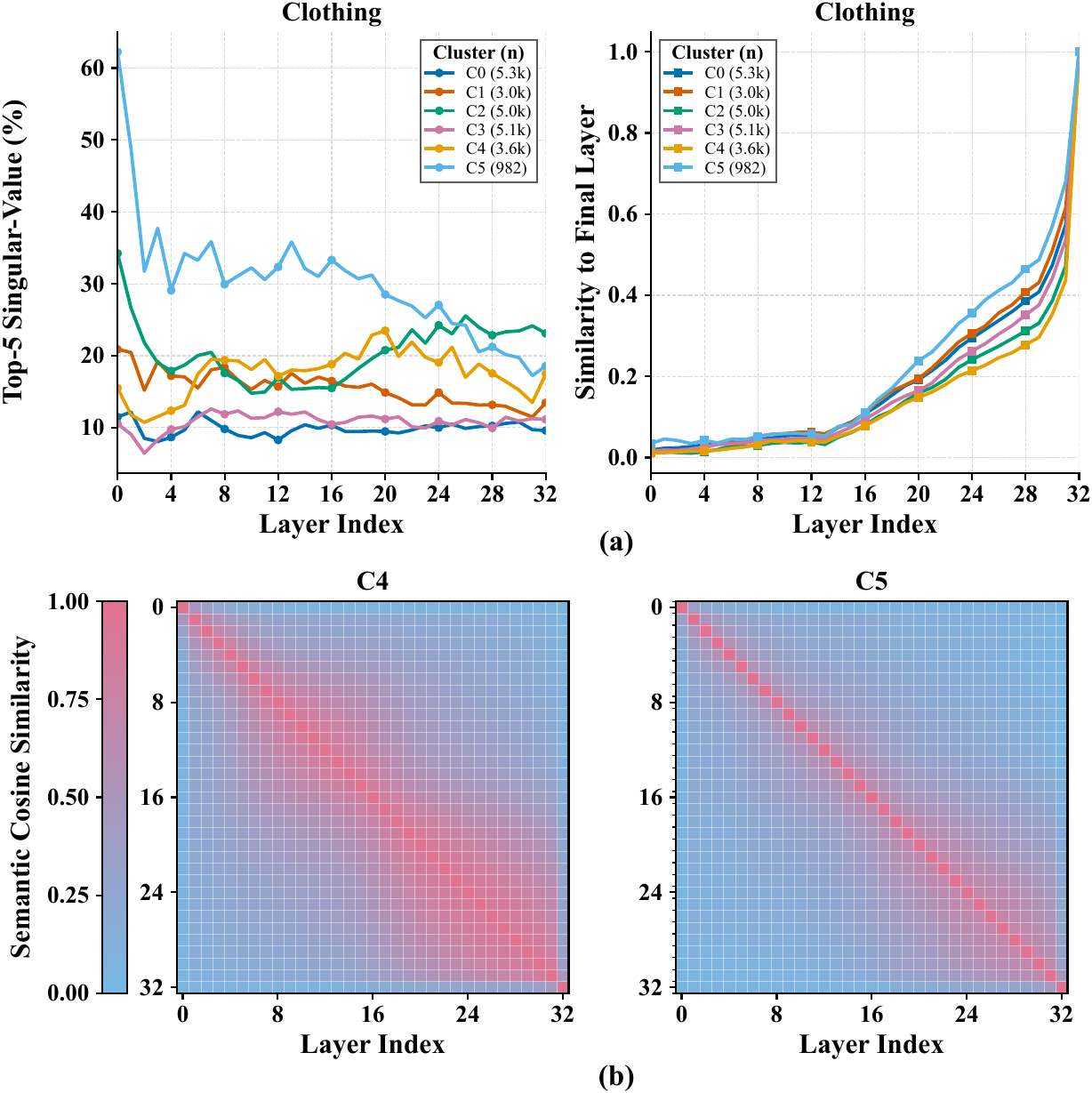}
    \vspace{-0.3cm}
    \caption{Different item semantic groups exhibit (a) distinct Top-5 singular-value proportions and cosine similarities to the final layer, and (b) different inter-layer similarity structures across LLM layers.}
    \label{fig:item-heterogeneity}
    \vspace{-0.5cm}
\end{figure}

\textbf{3) Different item types exhibit varying semantic patterns across LLM layers.}
We cluster items based on their semantic similarity and track how their representations evolve across layers (\cf Figure~\ref{fig:item-heterogeneity}). Different item groups show markedly divergent evolution patterns: some suffer from rapid spectral collapse in deeper layers, whereas others evolve smoothly across depths. This heterogeneity suggests that information from different layers should be exploited differently for different items.

These observations indicate that relying solely on the final-layer embeddings is suboptimal, as they may suffer from severe spectral collapse while overlooking the rich semantics inherent in intermediate layers. To address this issue, we propose \textbf{IMFuse}, a novel \underline{\textbf{I}}nstance-Aware \underline{\textbf{M}}ulti-Layer \underline{\textbf{Fusion}} strategy that adaptively aggregates multi-layer semantic information to enhance item representations. Specifically, inspired by Observations 1 and 2, IMFuse introduces an \emph{Adaptive Multi-Layer Fusion} module that aggregates semantic information from all LLM layers, using learnable layer- and dimension-wise fusion weights to capture the contributions of different layers. Inspired by Observation 3, IMFuse further introduces an \emph{Instance-Aware Mixture-of-Experts (MoE)} module, which conditions the fusion weights on item semantics, so that items with diverse semantic characteristics can adopt heterogeneous fusion strategies tailored to their own layer-wise patterns. Rather than learning a separate fusion matrix for each item, this module shares a few lightweight experts across items, keeping the additional parameters independent of the number of items.  By integrating semantic knowledge from multiple LLM layers with item-specific adaptation, IMFuse constructs more informative semantic embeddings for SR.

Notably, IMFuse also offers several practical advantages. First, it serves as a plug-and-play module that can be seamlessly integrated into various SR backbones and LLM-based enhancement methods without modifying their architectures, enriching their item embeddings and boosting their performance. Second, it introduces only a small number of additional trainable parameters and incurs limited computational overhead. Extensive experiments on four real-world datasets with two SR backbones and four semantic enhancement methods show that IMFuse yields consistent improvements, with an average relative improvement of 6.72\%.


\begin{itemize}[leftmargin=*]
  \item We reveal that exploiting multi-layer LLM embeddings is crucial for sequential recommendation: the commonly used final-layer embeddings suffer from severe spectral collapse, while intermediate layers provide complementary semantics.


  \item We propose IMFuse, a novel instance-aware multi-layer fusion strategy for sequential recommendation, which adaptively integrates the multi-layer semantics of LLMs while customizing the fusion for each item.


  \item Extensive experiments on four real-world datasets demonstrate that IMFuse consistently improves recommendation performance across different SR backbones and LLM-enhancement strategies.

\end{itemize}

\begin{figure*}[t]
  \centering
  \includegraphics[width=\textwidth]{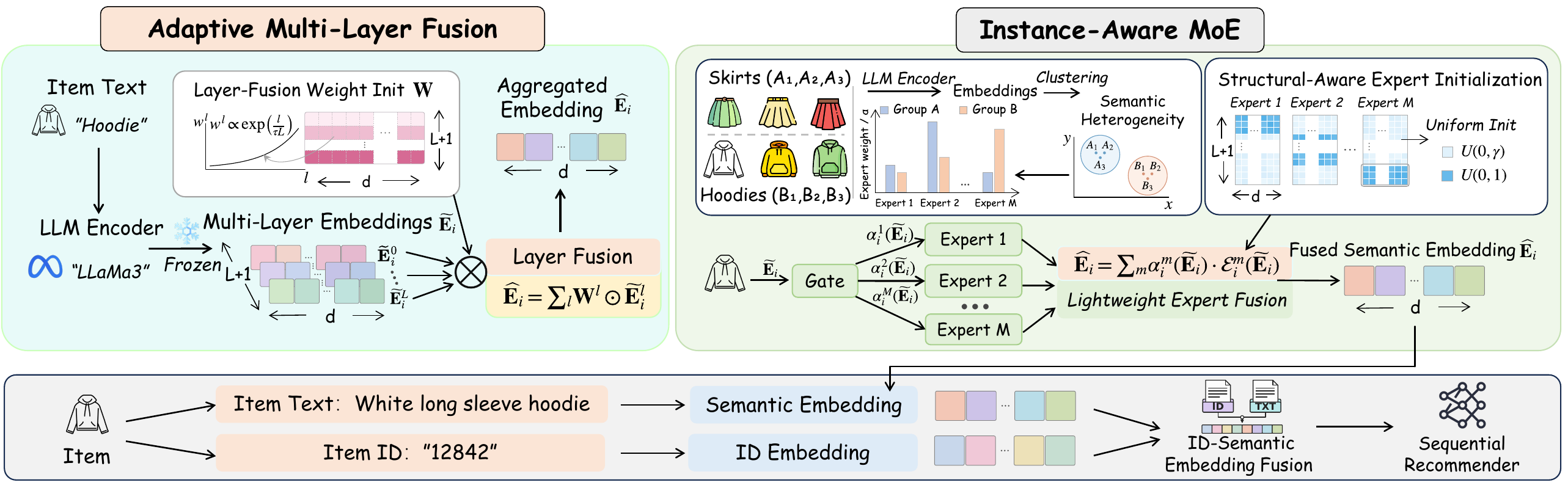}
  \caption{The overall framework of the proposed IMFuse.}
  \Description{The framework first extracts multiple representations from a frozen LLM and maps them into the recommendation space. It then combines shared dimension-wise layer preferences with item-specific expert routing to produce an item-specific multi-layer semantic representation, which replaces the conventional final-layer semantic input and is fed into the unchanged sequential recommender.}

  \label{fig:framework-overview}
  \vspace{-0.3cm}
\end{figure*}

\section{Preliminary}
\label{sec:preliminaries}

\subsection{Task Formulation}

Sequential recommendation predicts a user's next interaction from the
chronological sequence of previously interacted items
~\cite{hidasi2016gru4rec,kang2018self,sun2019bert4rec}.
Given a user set $\mathcal U$ and an item set $\mathcal I$, let
$N=|\mathcal I|$ denote the number of items. The interaction history
of user $u\in\mathcal U$ is organized chronologically as
$S_u=(i_1,i_2,\ldots,i_t)$, where $t$ is the sequence length.
The goal is to predict the next item $i_{t+1}$ given the user's behavior sequence $S_u$.

Traditional sequential recommenders represent each item with a trainable
embedding associated with its discrete ID
~\cite{kang2018self,xie2022cl4srec,yang2023generic}.
Sequential architectures, such as RNN-based~\cite{hidasi2016gru4rec} and Transformer-based models~\cite{kang2018self,sun2019bert4rec}, then use these embeddings
to model user interactions.
Although effective, these methods primarily rely on discrete IDs
and may underutilize textual information, such as item titles and
categories~\cite{hou2022unisrec}. Such information explicitly
describes item properties and provides complementary semantic
signals for recommendation.

\subsection{Language Embedding Enhancement}

Recent methods use LLM-derived semantic representations to
enhance sequential recommenders~\cite{zhang2025llminit,cui2026spectran}.
This process generally consists of the following steps: semantic encoding and semantic enhancement.

\subsubsection{Semantic Encoding}
\label{sec:semantic-encoding}

Item textual descriptions contain semantic information that can
be encoded into high-dimensional representations~\cite{hou2022unisrec,yuan2023morec}.
Let $\mathbf X=[x_i]_{i=1}^{N}$ denote the textual descriptions of all items,
where $x_i$ is the description of item $i$.
For an LLM with $L$ layers and hidden dimension $D$, we denote the item semantic embeddings extracted from the $L$-th LLM layer:
\begin{equation}
\mathbf{E}^{L} 
= \operatorname{LLMEnc}(\mathbf{X}),
\label{eq:item-language-embedding}
\end{equation}
where $\mathbf{E}^{L}\in\mathbb{R}^{N\times D}$ denotes the item embeddings extracted from the $L$-th LLM layer.



Existing LLM semantic enhancement methods typically use the final-layer
embedding matrix $\mathbf E^L$ to enrich conventional recommenders~\cite{liu2024llmesr,zhang2025llminit}.
However, relying on a single layer may limit the utilization of
semantic information across depths.
Our empirical studies reveal that final-layer representations may exhibit spectral concentration~\cite{cui2026spectran}, while different layers capture information
at different semantic granularities~\cite{jawahar2019bert,tenney2019bert,liu2019linguistic}.
These observations motivate the use of  semantic embeddings from other layers
as potential complementary information.

\subsubsection{Semantic Enhancement}
LLM representations typically have a much larger dimension $D$ than the ID embeddings $d$ used by recommendation models (\eg 4096 vs 32 ).
The semantic embeddings are further integrated with the item ID embeddings to obtain the final item representation $\mathbf{E}^{\mathrm{final}}$:
\begin{equation}
\widetilde{\mathbf{E}}^{L}=f_{\Phi}(\mathbf{E}^{L}), 
\end{equation}
\begin{equation}
\mathbf{E}^{\mathrm{final}}
=
g
\left(
\mathbf{E}^{\mathrm{id}},
\widetilde{\mathbf{E}}^{L}
\right),
\label{eq:item-enhancement}
\end{equation}
where $\widetilde{\mathbf{E}}^{L} \in \mathbb{R}^{N \times d}$ is the corresponding representation in the recommendation space, and $f_{\Phi}(\cdot)$ denotes the transformation operation. $\mathbf{E}^{\mathrm{id}}$ denotes the item ID embeddings, and $g(\cdot)$ is the combination function (\eg add or concatenate). The final item representation is then fed into the downstream sequential recommender to incorporate item-side semantic information and improve next-item prediction.

Recent studies typically project the semantic representation extracted from the final LLM layer into the recommendation space. Depending on the method, this projection is implemented through learnable adapter-based mappings~\cite{ren2024rlmrec,liu2024llmesr} or SVD-based spectral transformations~\cite{zhang2025llminit,cui2026spectran}. In contrast, IMFuse focuses on extracting more informative item semantic representations by leveraging multiple LLM layers. IMFuse is complementary to these existing strategies and can be integrated with them to further enhance the resulting representations.

\section{Methodology}
\label{sec:methodology}

In this section, we introduce IMFuse, a multi-layer semantic
fusion framework that integrates LLM representations through
the Adaptive Multi-Layer Fusion and the Instance-Aware MoE module.
The overall framework is illustrated in Figure~3.

\subsection{Motivation Analysis}
\label{sec:motivation-analysis}

Existing language embedding enhancement methods typically
rely on representations from the final LLM layer.
This choice may overlook complementary information preserved
at other depths. In particular, final-layer representations
may exhibit spectral collapse~\cite{cui2026spectran},
while different layers encode information at different
semantic granularities
~\cite{jawahar2019bert,tenney2019bert,liu2019linguistic}.
We investigate these properties through the following analyses, which provide empirical motivation for the design of our multi-layer fusion mechanism.

\textbf{Motivation 1: Item embeddings from deeper LLM layers suffer from increasingly severe spectral collapse.}
For the \(l\)-th LLM layer, let \(\mathbf{E}^{l}\) denote the encoded item semantic embedding matrix, where each row corresponds to the semantic representation of one item. To study the spectral collapse, we first apply singular value decomposition (SVD)~\cite{hoecker1996svd} to
$\mathbf E^l$:
\begin{equation}
\mathbf E^l
=\mathbf U^l\mathbf\Sigma^l(\mathbf V^l)^\top,
\label{eq:layer-svd}
\end{equation}
where $\mathbf U^l \in \mathbb R^{N\times r}$ and $\mathbf V^l \in \mathbb R^{D \times r}$ are the left and
right singular matrices, respectively. $\mathbf\Sigma^l = \mathrm{diag}(\sigma_1^l, \sigma_2^l, \cdots, \sigma_r^l)\in \mathbb R^{r\times r}$ is the singular value matrix ($\sigma_1^l\geq\cdots\geq\sigma_{r}^l\geq0$).
Following previous work~\cite{hu2025alphafuse, cui2026spectran}, we compute the proportion of the top-K singular values in the overall spectrum to study the spectral collapse:
\begin{equation}
\rho_k^l
=
\frac{\sum_{j=1}^{k}(\sigma_j^l)^2}
     {\sum_{j=1}^{r}(\sigma_j^l)^2},
\label{eq:cumulative-explained-variance}
\end{equation}
A higher value of $\rho_k^l$ indicates that the spectrum is concentrated in the top few spectral directions, \ie high-dimensional semantic embeddings degenerate into a narrow low-dimensional subspace, severely degrading the representational capacity and reducing information richness~\cite{jing2022dimensional,chen2024dimensional,cui2026spectran}.


As shown in Figure~\ref{fig:layer-motivation} (left), item embeddings from deeper LLM layers exhibit increasingly severe spectral collapse. Only a small portion of the dominant spectral information is retained, while other minor spectral signals are suppressed or diluted, limiting the information richness of the item representations \cite{chen2024dimensional,guo2024embedding}. 
Such a severe spectral collapse may restrict the recommendation model from effectively modeling   collaborative signals during training~\cite{lin2025popularity,cui2026specformer}. 
Therefore, we exploit item embeddings from
multiple LLM layers to capture complementary semantic information beyond
the final layer.

\textbf{Motivation 2: Different LLM layers encode complementary item semantic information.}
Building on the preceding spectral analysis, we next examine whether different LLM layers preserve complementary item semantics. We measure the average cosine similarity of the items' semantic embeddings at layers $i$ and $j$:
\begin{equation}
s_{i,j}
=
\frac{1}{N}\sum_{n=1}^{N}
\frac{(\mathbf e_n^i)^\top\mathbf e_n^j}
     {\lVert\mathbf e_n^i\rVert_2
      \lVert\mathbf e_n^j\rVert_2},
\label{eq:inter-layer-similarity}
\end{equation}
where $n$ denotes the $n$-th item, $s_{i,j}$ is the  cosine similarity between layers $i$ and $j$ for all items, and $\lVert\cdot\rVert_2$ denotes the Euclidean
norm.

As shown in Figure~\ref{fig:layer-motivation} (right), we observe that adjacent LLM
layers have higher similarity, while more distant layers
exhibit larger differences. 
This suggests that different layers of the LLM encode semantics of complementary information. It further motivates us to equip recommendation models with complementary semantic knowledge from different LLM layers for more flexible semantic integration and selection, effectively facilitating the semantic alignment during the embedding fusion process.

\textbf{Motivation 3: Different item types exhibit varying semantic patterns across LLM layers.}
The preceding analysis highlights the necessity of utilizing multi-layer embeddings. However, items in recommender systems exhibit high heterogeneity~\cite{sun2019bert4rec,zhai2024hstu}. In practice, items may exhibit varying preferences for semantic embeddings across LLM layers in the downstream recommenders.
To study the item-level semantic heterogeneity, we apply
K-means++~\cite{arthur2007kmeanspp} to items' semantic
embeddings to divide different item groups.
Additional details are provided in
Appendix~\ref{app:item-group-analysis}.


As shown in Figure~\ref{fig:item-heterogeneity}, different item groups exhibit distinct layer-wise spectral collapse and similarity patterns.
For example, some item groups show a rapid decrease in spectral collapse with depth, whereas others exhibit limited changes (\eg C5 vs C0). Moreover, different item groups display heterogeneous semantic similarities: some items experience drastic semantic shifts across layers, while others remain relatively smooth (\eg C5 vs C4).
These observations motivate us to employ an instance-aware multi-layer fusion strategy for more flexible and effective semantic enhancement.

\subsection{Adaptive Multi-Layer Fusion}
\label{sec:global-layer-preference}
As analyzed in Motivations 1 and 2, utilizing multi-layer embeddings yields greater benefits than relying primarily on the final-layer embeddings. It not only mitigates the potential spectral collapse limitation of the last layer, but also equips the recommendation model with layer-wise richer semantic knowledge. Therefore, IMFuse first introduces a  multi-layer fusion mechanism, enabling the recommendation model to dynamically learn the layer-wise weights, aggregating semantics from different layers. Formally, the multi-layer fusion process can be defined as:
\begin{equation}
\widehat{\mathbf E}
=
\operatorname{Fusion}
\bigl(
\widetilde{\mathbf E}^{0},
\widetilde{\mathbf E}^{1},
\ldots,
\widetilde{\mathbf E}^{L}
\bigr),
\label{eq:general-aggregation}
\end{equation}
where $\widehat{\mathbf E}\in\mathbb R^{N\times d}$ is the semantic fused embedding matrix and $\operatorname{Fusion}(\cdot)$ operator denotes the multi-layer fusion function.

\textbf{Efficient Multi-layer Fusion.} The $\operatorname{Fusion}(\cdot)$ operator can be implemented by various strategies (\eg a learnable weight vector, matrix or MLP network). In this work, we implement our Adaptive Multi-Layer Fusion as a learnable layer-aware and dimension-aware weight matrix, which can be defined as:
\begin{equation}
\widehat{\mathbf E}_i
=
\sum_{l=0}^{L}
\mathbf W^l\odot\widetilde{\mathbf E}^l_i,
\label{eq:weighted-aggregation}
\end{equation}
where $\odot$ is the Hadamard product, $\mathbf W^l \in\mathbb R^{d}$ is the weight vector for the $l$-th layer, and $\mathbf W\in\mathbb R^{(L+1) \times d}$ is the full multi-layer fusion matrix.
On the one hand, we introduce layer-specific learnable parameters to capture the importance of each layer $l$ and control the strength of its semantic contribution. On the other hand, rather than sharing a single weight across all dimensions within a layer, we assign dimension-wise weights to capture dimensional heterogeneity. This design is motivated by our observation (\cf Appendix~\ref{app:dimension-analysis}).




\textbf{Implementation Details.}
For initialization, we investigate different strategies (\eg uniform, random and other layer-wise initialization) with detailed comparisons (\cf Appendix~\ref{app:initialization}), and we adopt a layer-wise ascending strategy:
\begin{equation}
w^l \propto \exp\left(\frac{l}{\tau L}\right),
\end{equation}
where \(\tau\) controls the concentration of the initial weights across layers, assigning larger initial values to deeper layers. The fusion parameters are jointly optimized with the transformation parameters under the sequential recommendation objective, and the fusion operation can also be applied after SVD-based dimensionality reduction when semantic embeddings are first reduced.

\subsection{Instance-Aware MoE}
\label{sec:instance-aware-moe}

As analyzed in Motivation 3, different items exhibit varying semantic and collapse patterns across LLM layers. Consequently, it is essential to extend \Cref{eq:weighted-aggregation} to an instance-aware strategy. 
A naive strategy is to assign an independent weight matrix to each item. However, this introduces \(\mathcal{O}(NLd)\) trainable parameters, leading to unaffordable costs.


\textbf{Instance-Aware MoE.} 
To address this issue, we introduce an instance-aware MoE. Instead of assigning separate parameters to each item, we use multiple diverse experts to capture different layer-wise fusion patterns, allowing each item to adaptively select experts based on its semantics and derive its
fusion weights. Formally, the MoE is formulated as:
\begin{equation}
\widehat{\mathbf E}_i=\sum_{m=1}^{M}
{\alpha}_i^{m}(\widetilde{\mathbf E}_i) \cdot \mathbf{\mathcal E}_i^{m}(\widetilde{\mathbf E}_i),
\end{equation}
\begin{equation}
\alpha_i^m(\widetilde{\mathbf{E}}_i)
=
\left[
\operatorname{Softmax}
\left(
\mathbf{g}_i(\widetilde{\mathbf{E}}_i)
\right)
\right]_m,
\end{equation}
   where $\widehat{\mathbf E}_i$ is the final multi-layer fused embedding, $M$ is the number of experts ($M=3$ in our experiments), and $\alpha_i^m(\cdot)$ is the normalized weight for the  expert $m$. $\widetilde{\mathbf{E}}_i$ denotes the embedding matrix of item $i$ after dimensionality reduction, $\mathbf{g}_i(\cdot)\in\mathbb{R}^{M}$ is the gating network (a 2-layer MLP with GELU activation~\cite{hendrycks2016gelu}), and Softmax denotes normalization across experts. $\mathbf{\mathcal E}_i^{m}(\cdot)$ is the $m$-th expert network. 



\textbf{Lightweight Expert Network.} 
In conventional MoE paradigms, expert networks typically consist of multiple MLPs. However, deploying such complex experts in recommender systems incurs excessive computational overhead~\cite{hou2022unisrec}. Furthermore, it exacerbates critical issues such as routing load imbalance~\cite{shazeer2017outrageously} and expert starvation~\cite{zhou2022expertchoice}. 
To further boost efficiency and enhance training stability, we devise a simple yet highly efficient expert network:
\begin{equation}
\mathbf{\mathcal E}_i^{m}(\widetilde{\mathbf E}_i) = \sum_{l=0}^{L}
\mathbf{W}^{l,m} \odot \widetilde{\mathbf E}_i^{l},
\end{equation}
where $\mathbf{W}^{l,m}$ is the weight vector for layer $l$ from $\mathbf{W}^{m} \in \mathbb{R}^{(L+1) \times d}$. This design avoids costly MLP transformations, reducing computational overhead.

\textbf{Structural-Aware Initialization.} Inspired by the empirical observations (\cf Figure~\ref{fig:layer-motivation}), the spectral collapse and semantic similarity evolution of embeddings exhibit piecewise structural properties across LLM layers. Consequently, we expect each expert to focus on the weight distribution of a specific structural interval.

Specifically, we evenly partition all $L+1$ layers into $M$ segments to perfectly align with the $M$ experts. The chunk size is defined as $K = \lfloor \frac{L+1}{M} \rfloor$, and the structural-aware initialization is defined as:
\begin{equation}
w^{l,m} = \mathbf{\xi}^{l,m} \cdot w^{l},
\end{equation}
\begin{equation}
\mathbf{\xi}^{l,m} \sim 
\begin{cases} 
\mathcal{U}(0, 1), & m=\min\{\lfloor l/K\rfloor+1,M\} \\
\mathcal{U}(0, \gamma), & \text{otherwise}
\end{cases}
\end{equation}
where $\mathbf{\xi}^{l,m}$ is independently sampled from a uniform distribution $\mathcal{U}$ to enforce the $m$-th expert to focus on its corresponding layer-aware segment, $\gamma < 1$ is a scaling factor used to suppress the semantic signals outside the target interval. 


\subsection{Discussion}
\label{sec:discussion}

In summary, the proposed IMFuse offers several compelling advantages, which can be highlighted as follows:

\begin{itemize}[leftmargin=*]
    \item \textbf{Adaptive Fusion.} Unlike conventional methods that apply uniform weights or rely primarily on final-layer embeddings, IMFuse dynamically captures instance-aware multi-layer weights. Different semantic granularities across LLM layers are fully considered, effectively mitigating the last-layer spectral collapse and endowing the model with a stronger representation capacity.
    \item \textbf{Model-Agnostic Flexibility.} IMFuse serves as a universal, plug-and-play semantic enhancement module. The multi-layer fused semantic embedding $\widehat{\mathbf E}$ can be seamlessly integrated into various downstream recommendation models, without requiring any modifications to their backbone architectures.    


    \item \textbf{High Efficiency.} We analyze the additional space and time overhead introduced by IMFuse. The additional trainable parameters come from two components: the gate network, a two-layer MLP with parameter complexity $\mathcal{O}(dL+L^2)$, and the expert network comprises $M$ parameter groups with complexity approximately $\mathcal{O}(MLd)$. For time complexity, processing $N$ items therefore incurs an overall computational cost of $\mathcal{O}(NMLd)$. In comparison, the existing dimensionality-reduction network requires $\mathcal{O}(NDd)$ computation.    Although the proposed fusion module introduces additional overhead,    the efficiency study in Section~\ref{sec:efficiency} shows that this cost remains acceptable.

    

\end{itemize}

\section{Experiments}
\label{sec:experiments}
\setcounter{dbltopnumber}{2}
We aim to answer the following research questions:

\begin{itemize}[leftmargin=*]
  \item \textbf{RQ1: } How does IMFuse perform compared with existing semantic enhancement and multi-layer fusion methods?
  \item \textbf{RQ2: } What are the impacts of different components of IMFuse?
  \item \textbf{RQ3: } How about the efficiency of IMFuse?
  \item \textbf{RQ4: } What layer preferences does IMFuse learn?
\end{itemize}

\subsection{Experimental Settings}

\subsubsection{Datasets}
We closely refer to recent work~\cite{zhang2025llminit,cui2026spectran} and also adopt four widely used real-world  datasets: \textit{Toys and Games}, \textit{Beauty}, 
\textit{Clothing, Shoes and Jewelry}, and 
\textit{Office Products}
\footnote{\url{https://cseweb.ucsd.edu/~jmcauley/datasets/amazon/}} from the Amazon products~\cite{mcauley2015amazon}.
Following recent work, we adopt the conventional leave-one-out protocol~\cite{kang2018self}. Then we sort each user's interactions by timestamp, reserve the last interaction for testing and the penultimate interaction for validation, and use all remaining interactions for training. Dataset statistics are summarized in \Cref{tab:datasets}.

\begin{table}[t]
  \caption{Statistics of the datasets.}
  \vspace{-0.3cm}
  \label{tab:datasets}
  \centering
  \small
  \setlength{\tabcolsep}{3.5pt}
  \begin{tabular}{ccccc}
    \toprule
    \textbf{Dataset} & \textbf{Users} & \textbf{Items} & \textbf{Interactions} & \textbf{Density} \\
    \midrule
    \textbf{Clothing} & 39,230 & 22,948 & 266,481 & 0.0296\% \\
    \textbf{Beauty}   & 22,332 & 12,086 & 168,446 & 0.0624\% \\
    \textbf{Toy}      & 19,124 & 11,757 & 141,630 & 0.0630\% \\
    \textbf{Office}   &  4,895 &  2,414 &  41,462 & 0.3509\% \\
    \bottomrule
  \end{tabular}
  \vspace{-0.3cm}
\end{table}

\begin{table*}[t]
  \caption{Overall performance across four real-world datasets. "Avg. Impr." denotes the average relative improvement of IMFuse over the four base enhancement methods; "N" and "H" denote NDCG and HR, respectively.}
  \label{tab:overall}
  \centering
  \scalebox{0.72}{%
  \begin{tabular}{@{}cc*{16}{c}@{}}
    \toprule
      & & \multicolumn{4}{c}{\textbf{Clothing}} & \multicolumn{4}{c}{\textbf{Beauty}}
        & \multicolumn{4}{c}{\textbf{Toy}} & \multicolumn{4}{c}{\textbf{Office}} \\
    \cmidrule(lr){3-6}\cmidrule(lr){7-10}\cmidrule(lr){11-14}\cmidrule(lr){15-18}
    \multirow{-2}{*}{\textbf{Backbone}} & \multirow{-2}{*}{\textbf{Method}}
      & N@10 & N@20 & H@10 & H@20 & N@10 & N@20 & H@10 & H@20
      & N@10 & N@20 & H@10 & H@20 & N@10 & N@20 & H@10 & H@20 \\
    \midrule
      & RLMRec            & 0.0131 & 0.0175 & 0.0261 & 0.0435 & 0.0293 & 0.0385 & 0.0585 & 0.0936 & 0.0320 & 0.0404 & 0.0671 & 0.1006 & 0.0405 & 0.0540 & 0.0825 & 0.1356 \\
      & + IMFuse          & 0.0135 & 0.0179 & 0.0276 & 0.0453 & 0.0302 & 0.0393 & 0.0599 & 0.0962 & 0.0328 & 0.0413 & 0.0698 & 0.1035 & 0.0465 & 0.0611 & 0.0968 & 0.1549 \\
      \cmidrule(lr){2-18}
      & LLM-ESR           & 0.0098 & 0.0128 & 0.0188 & 0.0309 & 0.0212 & 0.0281 & 0.0420 & 0.0692 & 0.0214 & 0.0279 & 0.0417 & 0.0676 & 0.0235 & 0.0314 & 0.0486 & 0.0803 \\
      & + IMFuse          & 0.0107 & 0.0139 & 0.0212 & 0.0339 & 0.0239 & 0.0315 & 0.0467 & 0.0768 & 0.0236 & 0.0292 & 0.0459 & 0.0685 & 0.0295 & 0.0408 & 0.0621 & 0.1073 \\
      \cmidrule(lr){2-18}
      & LLMInit           & 0.0117 & 0.0156 & 0.0232 & 0.0387 & 0.0270 & 0.0345 & 0.0527 & 0.0828 & 0.0295 & 0.0363 & 0.0600 & 0.0872 & 0.0370 & 0.0506 & 0.0801 & 0.1340 \\
      & + IMFuse          & 0.0121 & 0.0160 & 0.0243 & 0.0398 & 0.0278 & 0.0351 & 0.0547 & 0.0837 & 0.0302 & 0.0369 & 0.0632 & 0.0898 & 0.0401 & 0.0538 & 0.0825 & 0.1375 \\
      \cmidrule(lr){2-18}
      & SpecTran          & 0.0163 & 0.0217 & 0.0354 & 0.0572 & 0.0324 & 0.0417 & 0.0702 & 0.1069 & 0.0376 & 0.0469 & 0.0821 & 0.1189 & 0.0467 & 0.0604 & 0.1015 & 0.1549 \\
    \rowcolor[HTML]{96FFFB}
      \cellcolor{white} & \cellcolor{white} + IMFuse & \textbf{0.0175} & \textbf{0.0232} & \textbf{0.0378} & \textbf{0.0605} & \textbf{0.0336} & \textbf{0.0430} & \textbf{0.0721} & \textbf{0.1098} & \textbf{0.0398} & \textbf{0.0488} & \textbf{0.0857} & \textbf{0.1219} & \textbf{0.0475} & \textbf{0.0619} & \textbf{0.1034} & \textbf{0.1608} \\
    \cmidrule(lr){2-18}
    \multirow{-9}{*}{\textbf{SASRec}} & {Avg Impr.}
      & \imprcell{5.75\%} & \imprcell{5.09\%} & \imprcell{7.51\%} & \imprcell{5.61\%}
      & \imprcell{5.62\%} & \imprcell{4.76\%} & \imprcell{5.02\%} & \imprcell{4.39\%}
      & \imprcell{5.25\%} & \imprcell{3.15\%} & \imprcell{5.95\%} & \imprcell{2.43\%}
      & \imprcell{12.61\%} & \imprcell{12.97\%} & \imprcell{12.49\%} & \imprcell{13.57\%} \\
    \cmidrule(lr){1-18}
      & RLMRec            & 0.0115 & 0.0157 & 0.0237 & 0.0403 & 0.0263 & 0.0344 & 0.0517 & 0.0840 & 0.0233 & 0.0305 & 0.0475 & 0.0761 & 0.0381 & 0.0496 & 0.0753 & 0.1213 \\
      & + IMFuse          & 0.0127 & 0.0171 & 0.0261 & 0.0435 & 0.0286 & 0.0367 & 0.0581 & 0.0903 & 0.0274 & 0.0354 & 0.0564 & 0.0882 & 0.0399 & 0.0512 & 0.0791 & 0.1244 \\
      \cmidrule(lr){2-18}
      & LLM-ESR           & 0.0113 & 0.0150 & 0.0228 & 0.0376 & 0.0215 & 0.0280 & 0.0445 & 0.0706 & 0.0223 & 0.0289 & 0.0441 & 0.0701 & 0.0277 & 0.0383 & 0.0576 & 0.0999 \\
      & + IMFuse          & 0.0119 & 0.0160 & 0.0244 & 0.0407 & 0.0234 & 0.0307 & 0.0476 & 0.0767 & 0.0239 & 0.0306 & 0.0474 & 0.0741 & 0.0291 & 0.0400 & 0.0601 & 0.1036 \\
      \cmidrule(lr){2-18}
      & LLMInit           & 0.0103 & 0.0135 & 0.0212 & 0.0340 & 0.0264 & 0.0326 & 0.0515 & 0.0762 & 0.0268 & 0.0329 & 0.0535 & 0.0775 & 0.0398 & 0.0511 & 0.0793 & 0.1242 \\
      & + IMFuse          & 0.0111 & 0.0147 & 0.0231 & 0.0374 & 0.0272 & 0.0338 & 0.0526 & 0.0789 & 0.0277 & 0.0335 & 0.0557 & 0.0789 & 0.0427 & 0.0540 & 0.0858 & 0.1312 \\
      \cmidrule(lr){2-18}
      & SpecTran          & 0.0164 & 0.0216 & 0.0361 & 0.0566 & 0.0316 & 0.0405 & 0.0701 & 0.1054 & 0.0361 & 0.0452 & 0.0779 & 0.1141 & 0.0421 & 0.0549 & 0.0895 & 0.1408 \\
    \rowcolor[HTML]{96FFFB}
      \cellcolor{white} & \cellcolor{white} + IMFuse & \textbf{0.0170} & \textbf{0.0223} & \textbf{0.0374} & \textbf{0.0584} & \textbf{0.0329} & \textbf{0.0418} & \textbf{0.0720} & \textbf{0.1076} & \textbf{0.0371} & \textbf{0.0468} & \textbf{0.0801} & \textbf{0.1188} & \textbf{0.0437} & \textbf{0.0575} & \textbf{0.0968} & \textbf{0.1520} \\
    \cmidrule(lr){2-18}
   \multirow{-9}{*}{\textbf{HSTU}} & {Avg Impr.}
      & \imprcell{6.79\%} & \imprcell{6.93\%} & \imprcell{7.43\%} & \imprcell{7.34\%}
      & \imprcell{6.18\%} & \imprcell{5.80\%} & \imprcell{6.05\%} & \imprcell{5.44\%}
      & \imprcell{7.72\%} & \imprcell{6.83\%} & \imprcell{8.29\%} & \imprcell{6.88\%}
      & \imprcell{5.22\%} & \imprcell{4.52\%} & \imprcell{6.43\%} & \imprcell{4.96\%}\\
    \bottomrule
  \end{tabular}
  }
  \vspace{-0.3cm}
\end{table*}

\begin{table}[t]
  \caption{Comparison with multi-layer fusion baselines on Clothing and Toy.}
  \vspace{-0.3cm}
  \label{tab:fusion-comparison}
  \centering
  \resizebox{\columnwidth}{!}{%
  \begin{tabular}{@{}cc*{8}{c}@{}}
    \toprule
      & & \multicolumn{4}{c}{\textbf{Clothing}} & \multicolumn{4}{c}{\textbf{Toy}} \\
    \cmidrule(lr){3-6}\cmidrule(lr){7-10}
      & & \multicolumn{2}{c}{\textbf{RLMRec}} & \multicolumn{2}{c}{\textbf{SpecTran}}
        & \multicolumn{2}{c}{\textbf{RLMRec}} & \multicolumn{2}{c}{\textbf{SpecTran}} \\
    \cmidrule(lr){3-4}\cmidrule(lr){5-6}\cmidrule(lr){7-8}\cmidrule(lr){9-10}
    \multirow{-3}{*}{\textbf{Backbone}} & \multirow{-3}{*}{\textbf{Method}}
      & N@20 & H@20 & N@20 & H@20 & N@20 & H@20 & N@20 & H@20 \\
    \midrule
      & LAEF     & 0.0168 & 0.0423 & 0.0223 & 0.0573 & 0.0364 & 0.0908 & 0.0460 & 0.1166 \\
      & CASE-MLP & 0.0173 & 0.0436 & 0.0221 & 0.0576 & 0.0397 & 0.0969 & 0.0463 & 0.1185 \\
      & VA-HS    & 0.0171 & 0.0421 & 0.0220 & 0.0571 & 0.0379 & 0.0922 & 0.0451 & 0.1148 \\
    \rowcolor[HTML]{96FFFB}
      \cellcolor{white} & \cellcolor{white} IMFuse
        & \textbf{0.0179}
        & \textbf{0.0453}
        & \textbf{0.0232}
        & \textbf{0.0605}
        & \textbf{0.0413}
        & \textbf{0.1035}
        & \textbf{0.0488}
        & \textbf{0.1219} \\
    \cmidrule(lr){2-10}
    \multirow{-5}{*}{\textbf{SASRec}} & {Impr.}
      & \imprcell{3.47\%} & \imprcell{3.90\%} & \imprcell{4.04\%} & \imprcell{5.03\%}
      & \imprcell{4.03\%} & \imprcell{6.81\%} & \imprcell{5.40\%} & \imprcell{2.87\%}\\    \cmidrule(lr){1-10}
      & LAEF     & 0.0167 & 0.0423 & 0.0214 & 0.0556 & 0.0322 & 0.0829 & 0.0426 & 0.1076 \\
      & CASE-MLP & 0.0165 & 0.0415 & 0.0216 & 0.0565 & 0.0337 & 0.0854 & 0.0451 & 0.1148 \\
      & VA-HS    & 0.0164 & 0.0413 & 0.0215 & 0.0567 & 0.0326 & 0.0818 & 0.0426 & 0.1081 \\
    \rowcolor[HTML]{96FFFB}
      \cellcolor{white} & \cellcolor{white} IMFuse
        & \textbf{0.0171}
        & \textbf{0.0435}
        & \textbf{0.0223}
        & \textbf{0.0584}
        & \textbf{0.0354}
        & \textbf{0.0882}
        & \textbf{0.0468}
        & \textbf{0.1188} \\
    \cmidrule(lr){2-10}
    \multirow{-5}{*}{\textbf{HSTU}} & {Impr.}
      & \imprcell{2.40\%} & \imprcell{2.84\%} & \imprcell{3.24\%} & \imprcell{3.00\%}
      & \imprcell{5.04\%} & \imprcell{3.28\%} & \imprcell{3.77\%} & \imprcell{3.48\%} \\\bottomrule
  \end{tabular}
  }
  \vspace{-0.3cm}
\end{table}

\begin{table*}[t]
  \caption{Ablation study of IMFuse with SpecTran.}
  \vspace{-0.3cm}
  \label{tab:ablation}
  \centering
  \resizebox{\textwidth}{!}{%
  \begin{tabular}{@{}cc*{16}{c}@{}}
    \toprule
      & & \multicolumn{4}{c}{\textbf{Clothing}} & \multicolumn{4}{c}{\textbf{Beauty}}
        & \multicolumn{4}{c}{\textbf{Toy}} & \multicolumn{4}{c}{\textbf{Office}} \\
    \cmidrule(lr){3-6}\cmidrule(lr){7-10}\cmidrule(lr){11-14}\cmidrule(lr){15-18}
    \multirow{-2}{*}{\textbf{Backbone}} & \multirow{-2}{*}{\textbf{Variant}}
      & N@10 & N@20 & H@10 & H@20 & N@10 & N@20 & H@10 & H@20
      & N@10 & N@20 & H@10 & H@20 & N@10 & N@20 & H@10 & H@20 \\
    \midrule
      & \emph{w/ Last.} & 0.0163 & 0.0217 & 0.0354 & 0.0572 & 0.0324 & 0.0417 & 0.0702 & 0.1069 & 0.0376 & 0.0469 & 0.0821 & 0.1189 & 0.0467 & 0.0604 & 0.1015 & 0.1549 \\
      & \emph{w/ IM.} & 0.0165 & 0.0217 & 0.0372 & 0.0580 & 0.0327 & 0.0419 & 0.0718 & 0.1091 & 0.0390 & 0.0480 & 0.0855 & 0.1214 & 0.0473 & 0.0608 & 0.1028 & 0.1524 \\
      & \emph{w/o M.} & 0.0170 & 0.0228 & 0.0372 & 0.0604 & 0.0326 & 0.0421 & 0.0710 & 0.1089 & 0.0384 & 0.0478 & 0.0838 & 0.1210 & 0.0472 & 0.0613 & 0.1021 & 0.1553 \\
      & \emph{w/ Rand.} & 0.0161 & 0.0215 & 0.0362 & 0.0576 & 0.0321 & 0.0417 & 0.0719 & 0.1097 & 0.0373 & 0.0466 & 0.0817 & 0.1186 & 0.0460 & 0.0604 & 0.0989 & 0.1561 \\
      & \emph{w/ Perm.} & 0.0163 & 0.0214 & 0.0356 & 0.0560 & 0.0319 & 0.0416 & 0.0712 & 0.1096 & 0.0375 & 0.0467 & 0.0824 & 0.1188 & 0.0463 & 0.0605 & 0.0993 & 0.1559 \\
      \cmidrule(lr){2-18}
    \rowcolor[HTML]{96FFFB}
    \cellcolor{white}\multirow{-6}{*}{\textbf{SASRec}} & \cellcolor{white} IMFuse
      & \textbf{0.0175} & \textbf{0.0232}
      & \textbf{0.0378} & \textbf{0.0605}
      & \textbf{0.0336} & \textbf{0.0430}
      & \textbf{0.0721} & \textbf{0.1098}
      & \textbf{0.0398} & \textbf{0.0488}
      & \textbf{0.0857} & \textbf{0.1219}
      & \textbf{0.0475} & \textbf{0.0619}
      & \textbf{0.1034} & \textbf{0.1608} \\
    \cmidrule(lr){1-18}
      & \emph{w/ Last.} & 0.0164 & 0.0216 & 0.0361 & 0.0566 & 0.0316 & 0.0405 & 0.0701 & 0.1054 & 0.0361 & 0.0452 & 0.0779 & 0.1141 & 0.0421 & 0.0549 & 0.0895 & 0.1408 \\
      & \emph{w/ IM.} & 0.0164 & 0.0213 & 0.0347 & 0.0544 & 0.0320 & 0.0410 & 0.0697 & 0.1055 & 0.0368 & 0.0460 & 0.0796 & 0.1173 & 0.0429 & 0.0559 & 0.0940 & 0.1459 \\
      & \emph{w/o M.} & 0.0169 & 0.0222 & 0.0365 & 0.0578 & 0.0321 & 0.0410 & 0.0696 & 0.1052 & 0.0369 & 0.0459 & 0.0798 & 0.1167 & 0.0428 & 0.0562 & 0.0938 & 0.1479 \\
      & \emph{w/ Rand.} & 0.0164 & 0.0215 & 0.0347 & 0.0551 & 0.0315 & 0.0399 & 0.0699 & 0.1036 & 0.0358 & 0.0453 & 0.0785 & 0.1160 & 0.0414 & 0.0542 & 0.0883 & 0.1397 \\
      & \emph{w/ Perm.} & 0.0163 & 0.0217 & 0.0355 & 0.0570 & 0.0312 & 0.0401 & 0.0691 & 0.1042 & 0.0358 & 0.0448 & 0.0782 & 0.1141 & 0.0411 & 0.0543 & 0.0880 & 0.1406 \\
      \cmidrule(lr){2-18}
    \rowcolor[HTML]{96FFFB}
    \cellcolor{white}\multirow{-6}{*}{\textbf{HSTU}} & \cellcolor{white} IMFuse
      & \textbf{0.0170} & \textbf{0.0223}
      & \textbf{0.0374} & \textbf{0.0584}
      & \textbf{0.0329} & \textbf{0.0418}
      & \textbf{0.0720} & \textbf{0.1076}
      & \textbf{0.0371} & \textbf{0.0468}
      & \textbf{0.0801} & \textbf{0.1188}
      & \textbf{0.0437} & \textbf{0.0575}
      & \textbf{0.0968} & \textbf{0.1520} \\
    \bottomrule
  \end{tabular}
  }
  \vspace{-0.3cm}
\end{table*}

\begin{table}
\centering
\caption{Efficiency study of IMFuse.}
\vspace{-0.3cm}
\label{tab:efficiency}
\scalebox{0.8}{
\begin{tabular}{cccc}
\toprule
\textbf{Method} & \textbf{Trainable   Parameters} & \textbf{Training Cost} & \textbf{Inference Cost} \\ \midrule

RLMRec & 9.61M & 1.92s & 1.00s \\
LLM-ESR & 9.53M & 2.65s & 1.07s \\
SpecTran & 2.21M & 1.73s & 0.67s \\
\midrule
SpecTran + IMFuse & 2.24M & 1.79s & 0.72s \\
 \bottomrule
\end{tabular}
}
\end{table}

\subsubsection{Backbones and Baselines} 
We use two well-known and representative sequential recommendation models as the backbones: \textbf{SASRec (ICDM'18)~\cite{kang2018self}}
uses causal self-attention for next-item prediction, while
\textbf{HSTU (ICML'24)~\cite{zhai2024hstu}}
adopts pointwise aggregated attention with time-aware relative bias.

Meanwhile, IMFuse can also be integrated with existing LLM-based semantic enhancement strategies. We implement and evaluate the following methods: \textbf{RLMRec (WWW'24)~\cite{ren2024rlmrec}} aligns semantic and ID
embeddings via MLP reconstruction;
\textbf{LLM-ESR (NeurIPS'24)~\cite{liu2024llmesr}} projects frozen LLM
embeddings into recommendation space and jointly models semantic and
collaborative views;
\textbf{LLMInit (EMNLP'25)~\cite{zhang2025llminit}} selects
high-variance semantic dimensions for ID embedding initialization; and
\textbf{SpecTran (SIGIR'26)~\cite{cui2026spectran}} uses a
spectral-aware Transformer adapter for LLM representation transformation.


Some studies in other domains have explored multi-layer fusion. We adapt these methods to our setting for comparison, including: \textbf{LAEF (arXiv'25)~\cite{gwak2025layeraware}} applies average, max,
and min aggregation over selected layers;
\textbf{CASE-MLP (EACL'26)~\cite{zhang2026case}} uses supervised MLP
projection for multi-layer fusion; and
\textbf{VA-HS (ICML'26)~\cite{zhang2026attentionvalues}} adopts
cross-layer value aggregation on cached hidden states.

\subsubsection{Evaluation Metrics}
Following  previous
work~\cite{zhang2025llminit}, we evaluate  next-item
ranking using Hit Ratio (HR@$K$)~\cite{kang2018self} and Normalized Discounted Cumulative Gain (NDCG@$K$)~\cite{jarvelin2002ndcg}, with $K\in\{10,20\}$.
HR evaluates whether a relevant item appears in the top-$K$ recommendations.
NDCG evaluates recommendation quality accounting for both relevance scores
and their positions.

\subsubsection{Implementation Details}
Following previous work, we use Adam~\cite{kingma2015adam} optimizer with a learning rate of $\{10^{-5},10^{-4},10^{-3}\}$, weight decay over $\{0,10^{-6},10^{-5},10^{-4}\}$, and dropout from 0 to 0.5 with a step size of 0.1. The batch size is 256, the maximum history length is 10~\cite{yang2023generic}, and the recommendation dimension is $d=128$; SASRec uses two Transformer blocks and one attention head, while HSTU uses only the ID features of items to fit the setting of sequential recommendation. Unless otherwise specified, we use LLaMA-3-8B~\cite{grattafiori2024llama} as the LLM to encode item titles for semantic embeddings. 
For our method,  we select $\tau$ from 0.2 to 1 with the step of 0.2 and $\gamma$ over $\{0.3,0.5\}$ in practice. The hidden dimension of the two-layer gate MLP is $h=33$. The maximum number of epochs is 200, and we apply early stopping~\cite{prechelt1998early} with a patience of 10
based on validation NDCG@20. For all baselines, we follow the hyperparameter settings in their original papers and perform careful tuning to obtain the best-performing configurations in our search.
Models are implemented in PyTorch~\cite{paszke2019pytorch} and run on 4 NVIDIA RTX 5090 GPUs. Our code is available here.

\subsection{Performance Evaluation (RQ1)}

\Cref{tab:overall,tab:fusion-comparison} present the performance comparison between IMFuse and the baseline methods. We observe that:

\textbf{1) Overall performance comparisons.}
Across all evaluated configurations, IMFuse consistently improves the
corresponding base models. Specifically, IMFuse achieves
average relative improvements of 7.01\% and 6.43\% on SASRec and HSTU. These results demonstrate the effectiveness of IMFuse
in exploiting hierarchical LLM representations for recommendation.
Moreover, IMFuse can be integrated into existing semantic
enhancement pipelines without modifying their original mechanisms.

\textbf{2) Compared with final-layer enhancement baselines.}
Most existing language embedding enhanced recommendation methods rely on a single
final-layer LLM embedding, while these methods
mainly focus on transforming or selecting a fixed semantic representation,
but do not explicitly model how different layers contribute to different
items. We compare IMFuse with RLMRec~\cite{ren2024rlmrec},
LLM-ESR~\cite{liu2024llmesr}, LLMInit~\cite{zhang2025llminit}, and
SpecTran~\cite{cui2026spectran}, which improve final-layer representations through alignment,
dimension selection, or spectral adaptation. As shown in
\Cref{tab:overall}, IMFuse consistently improves all final-layer-based
baselines across different datasets and backbones, demonstrating the
benefit of adaptive multi-layer fusion and instance-aware MoE module. In contrast, IMFuse learns shared layer contribution patterns and
further adapts them according to item characteristics, enabling more
flexible utilization of hierarchical LLM representations.

\textbf{3) Compared with multi-layer fusion baselines.}
Beyond final-layer enhancement methods, we further evaluate IMFuse
against multi-layer fusion approaches, including LAEF~\cite{gwak2025layeraware},
CASE-MLP~\cite{zhang2026case},
and VA-HS~\cite{zhang2026attentionvalues}.
These methods generally aggregate hidden states through predefined fusion
strategies, such as averaging, projection, or attention, without
explicitly modeling recommendation-specific preference signals and
item-level layer heterogeneity. As shown in
Table~\ref{tab:fusion-comparison}, IMFuse achieves the best performance
in all 16 comparisons, with an average relative improvement of 3.91\%.
The results demonstrate that recommendation-oriented multi-layer fusion
requires adaptive modeling of layer contributions rather than simple
aggregation.

\subsection{Ablation Study (RQ2)}
As shown in \Cref{tab:ablation}, we perform the following ablation study to investigate the effects of each component:
1) \emph{w/ Last.}:  directly use the transformed final-layer representation $\widetilde{\mathbf E}^{L}$;
2) \emph{w/ IM.}:  initialize the layer-fusion weights with their mean value across layers;
3) \emph{w/o M.}:  remove instance-aware MoE module;
4) \emph{w/ Rand.}:  replace the depth-biased initialization of the shared layer matrix with
Gaussian initialization;
5) \emph{w/ Perm.}:  randomly permute the layer positions within each expert’s initialization while keeping the LLM layer representations unchanged. The permuted experts are then trained normally. We observe that:

1) Specifically, \emph{w/ Last.} demonstrates that intermediate layers contain complementary information beyond the final layer. Moreover, the
inferior performance of \emph{w/ IM.} indicates that mean
initialization provides a less effective starting point than the proposed depth-biased initialization. These results validate the necessity of adaptive multi-layer fusion for exploiting hierarchical LLM representations. 2) Removing instance-aware MoE module (\emph{w/o M.}) consistently
degrades recommendation performance. This verifies that adaptive multi-layer fusion alone cannot capture the heterogeneous semantic
characteristics across items. By introducing instance-aware MoE module,
IMFuse can adjust layer weights according to individual item
properties. 3) Different initialization strategies also affect the performance.
Compared with random initialization and permuted layer-preference
experts, IMFuse achieves better results, demonstrating the
effectiveness of incorporating depth-biased initialization and the MoE module. These results indicate that appropriate initialization
provides a more effective starting point for learning layer weights
and improves the optimization of adaptive multi-layer fusion.

\subsection{Efficiency Study (RQ3)}
\label{sec:efficiency}
\Cref{tab:efficiency} compares different semantic enhancement methods on the Toy dataset with the SASRec backbone. We report the number of trainable parameters, training time, and inference time.

The expert network accounts for most of the additional trainable parameters. In our setting, the number of experts is $M=3$, much smaller than the number of items $N$. IMFuse adds approximately 0.03M trainable parameters, far fewer than the backbone's approximately 2.21M parameters, representing an increase of only 1.36\%. Although IMFuse introduces additional computation, its cost is small relative to that of the dimensionality-reduction network. For SASRec with SpecTran, IMFuse adds approximately 0.06\,s to the backbone training time of 1.73\,s, corresponding to an overhead of 3.47\%. Inference time also increases modestly, by 7.46\%, while NDCG@20 improves by 4.05\%. Together with the average performance improvement of 6.72\% across all evaluated settings, these results demonstrate a favorable trade-off between recommendation effectiveness and computational overhead.

\subsection{Case Study (RQ4)}



To analyze the fusion behavior learned under recommendation supervision, we visualize learned layer weights and item-specific routing patterns.

\begin{figure}[t]
  \centering

  \includegraphics[width=\columnwidth]{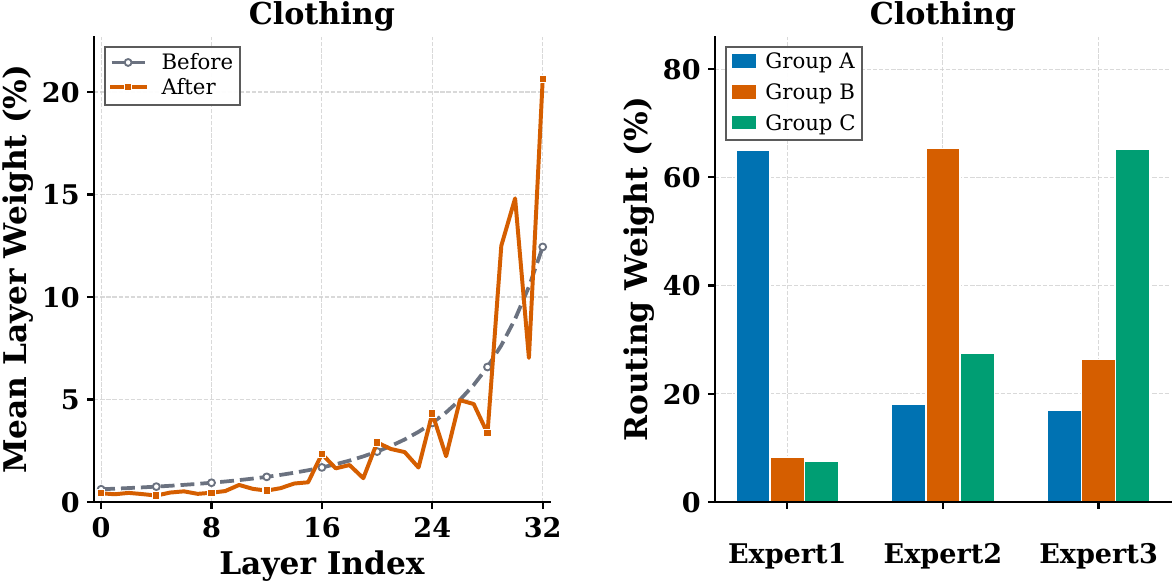}

  \vspace{-0.3cm}
  \caption{The layer-weight evolution and expert-routing patterns across
representative item groups with LLaMA-3-8B as the LLM backbone. }
  \vspace{-0.3cm}

  \label{fig:routing-analysis}
  
\end{figure}

As shown in \Cref{fig:routing-analysis} (left), we compare the layer-weight distributions before and after training. The learned distribution deviates substantially from its initialization, showing that the model adaptively adjusts layer weights under recommendation supervision. Moreover, the learned profile selectively emphasizes different layer representations across depth, indicating that useful semantic information is not restricted to a fixed layer range.

\Cref{fig:routing-analysis} (right) further reveals different expert contributions across item groups. Different groups exhibit distinct preferences over experts, suggesting that IMFuse learns item-specific expert routing weights. Each expert receives substantial weight from at least one group, with no single expert dominating all groups. This suggests that routing has not collapsed to a single expert in the visualized groups, consistent with the design motivation of the Lightweight Expert Network and Structural-Aware Initialization. Together with the ablation results in \Cref{tab:ablation}, these observations provide evidence that both Adaptive Multi-Layer Fusion and Instance-Aware MoE module contribute to the effectiveness of IMFuse.

\section{Related Work}
\label{sec:related-work}

\subsection{Sequential Recommendation}
Sequential recommendation predicts the next item from chronologically ordered historical interactions~\cite{kang2018self,xie2022cl4srec}. Early methods used recurrent or convolutional networks: GRU4Rec models within-session dependencies using gated recurrent units~\cite{hidasi2016gru4rec}, whereas Caser extracts local sequential patterns with convolutions~\cite{tang2018caser}. Self-attention later became a mainstream paradigm. SASRec uses causal self-attention~\cite{kang2018self}, BERT4Rec learns with bidirectional attention and masked-item prediction~\cite{sun2019bert4rec}, and HSTU introduces hierarchical sequential transduction units for large-scale generative recommendation~\cite{zhai2024hstu}. However, these methods mainly rely on interaction signals and overlook
rich item-side semantics. Recent LLM-enhanced recommendation methods
address this limitation by using LLMs to encode item texts into semantic
representations.

\subsection{LLMs for Recommendation}

Large language models can act directly as recommenders or serve as semantic encoders~\cite{zhao2024survey}. We next introduce these two paradigms.

\subsubsection{LLM-enhanced recommenders}
This paradigm primarily leverages the rich knowledge and reasoning capabilities of LLMs to enhance traditional recommender models~\cite{liu2024once,sun2024llmcf,ren2024aligned}. 

\textbf{Semantic Embedding Enhancement}. Offline semantic enhancement methods map pretrained content representations into a collaborative representation space~\cite{du2024disco,sheng2025language,he2025llm2rec}. Earlier methods use modality or content features: MoRec maps pretrained modality features~\cite{yuan2023morec}, and UniSRec uses a Mixture-of-Experts adapter for cross-domain sequence representations~\cite{hou2022unisrec}. LLM-based methods include RLMRec, which aligns language and collaborative representations through reconstruction or cross-view objectives~\cite{ren2024rlmrec}, and LLM-ESR, which jointly models semantic and collaborative views for long-tail recommendation~\cite{liu2024llmesr}. Geometry-oriented methods include WhitenRec, LLMInit, and SpecTran~\cite{zhang2024whitenrec,zhang2025llminit,cui2026spectran}. These methods primarily transform, align, or fuse a predetermined language embedding, while paying limited attention to which layer representations should constitute that embedding~\cite{hu2025alphafuse,lin2024bridging}.

\textbf{Other Enhancement Methods}. Beyond semantic embedding transformation, LLMRec uses LLM-based graph augmentation to improve recommendation data~\cite{wei2024llmrec}, whereas LLM4DSR uses an LLM to identify and denoise potentially noisy sequential interactions~\cite{wang2025llm4dsr}. Knowledge-distillation methods reduce the deployment burden of large models: DLLM2Rec transfers student-friendly knowledge to a conventional sequential recommender~\cite{cui2024distillation}, while SLIM distills a larger LLM into a smaller reasoning model for recommendation~\cite{wang2024can}. These methods enhance recommendation from data or model-transfer perspectives and are complementary to our focus on multi-layer item representations~\cite{xi2024openworld,jia2025learn,xu2025slmrec}.

\subsubsection{LLM-based Recommenders}

Another line directly uses a pretrained LLM as the recommender backbone~\cite{lin2024rella,shi2024planners,wang2025msl}, including unified frameworks that formulate recommendation as language processing and instruction-tuned systems~\cite{geng2022p5,bao2023tallrec,liao2024llara}. Early work also explored zero-shot next-item ranking by prompting general-purpose LLMs~\cite{hou2024zeroshot}. Subsequent studies fine-tune LLMs with recommendation data to reduce the mismatch between language modeling and recommendation~\cite{kim2024llmcf,tan2024idgenrec,zheng2024collaborative}, or organize user and item modeling hierarchically~\cite{chen2024hllm}. This line is outside our scope; a broader taxonomy is available in the survey~\cite{zhao2024survey}.

\subsection{Multi-Layer Language Embedding Fusion}

Although language embedding fusion has been explored in recent LLM-related studies, these methods do not account for the characteristics of recommendation.
For example, SBERT-WK~\cite{wang2020sbertwk} constructs sentence embeddings based on inter-layer geometry, but it does not learn layer contributions from user--item interactions.
Holistic Sentence Embeddings~\cite{chen2022holistic} aggregates information across tokens and layers, but it does not consider item-dependent layer preferences in recommendation.
Pooling and Attention~\cite{tang2024pooling} investigates trainable pooling strategies for LLM-based embeddings, but its aggregation is not guided by next-item prediction or collaborative signals.
Layer-Aware Embedding Fusion~\cite{gwak2025layeraware} compares different layer aggregation strategies and layer ranges for text classification, but it does not model the interaction between multi-layer item semantics and collaborative item-ID embeddings.
In contrast, IMFuse explicitly learns dimension-wise global layer preferences and instance-aware expert modulation under the sequential recommendation objective, enabling adaptive layer contributions for different dimensions and items and leading to improved recommendation performance.

\section{Conclusion}
\label{sec:conclusion}

This paper revisits the common practice of enhancing sequential recommenders using only final-layer LLM embeddings and demonstrates that intermediate-layer representations can provide complementary semantic signals, while final-layer representations may suffer from spectral collapse. To exploit such hierarchical information, we propose IMFuse, an instance-aware multi-layer fusion module that learns layer weights and dynamically adapts them for each item through an instance-aware MoE module. Extensive experiments on four real-world datasets demonstrate that our method consistently improves recommendation performance across different SR backbones and semantic enhancement methods, with only modest parameter and computational overhead. Future work will explore more memory-efficient strategies to reduce the storage and computational costs of multi-layer representations.

\bibliographystyle{ACM-Reference-Format}
\bibliography{references}
\clearpage
\appendix
\section{Additional Experimental Analysis}
\label{app:additional-analysis}

\subsection{Layer-wise Representation and Initialization Analysis}
\label{app:initialization}

To motivate Adaptive Multi-Layer Fusion, we examine how layer depth and fusion strategies affect recommendation performance.
Specifically, we try to evaluate individual layer
representations on representative datasets and different backbones. These exploratory experiments were conducted with preliminary configurations across different semantic enhancement frameworks, which differ from those used in the main experiments. As shown in
\Cref{tab:layer-representation}, single-layer performance generally increases
with layer depth, indicating that deeper representations provide stronger
standalone recommendation signals.

However, stronger final-layer performance does not imply that intermediate
layers are redundant. As shown in \Cref{tab:multi-layer-fusion}, simple
multi-layer fusion can further improve over the final-layer baseline in
some settings, suggesting that intermediate layers preserve complementary
recommendation-relevant information and motivating adaptive multi-layer
fusion.

\begin{table}[!htbp]
\centering
\caption{Comparison of representations from different LLM layers on Beauty and Clothing.}
\label{tab:layer-representation}

\begin{tabular}{@{}lcccc@{}}
\toprule
\multirow{2}{*}{\textbf{Embedding}}
& \multicolumn{2}{c}{\textbf{Beauty}}
& \multicolumn{2}{c}{\textbf{Clothing}} \\
\cmidrule(lr){2-3}\cmidrule(l){4-5}
& \textbf{N@20} & \textbf{H@20}
& \textbf{N@20} & \textbf{H@20} \\
\midrule
Layer 08         & 0.0321 & 0.0779 & 0.0184 & 0.0462 \\
Layer 16         & 0.0327 & 0.0806 & 0.0200 & 0.0521 \\
Layer 24         & 0.0334 & 0.0816 & 0.0207 & 0.0527 \\
\rowcolor[HTML]{96FFFB}
Layer 32 (Final) & \textbf{0.0390} & \textbf{0.0989}
                 & \textbf{0.0214} & \textbf{0.0542} \\
\bottomrule
\end{tabular}
\end{table}

\begin{table}[!htbp]
\centering
\caption{Comparison of multi-layer fusion strategies on Beauty and Clothing.}
\label{tab:multi-layer-fusion}
\small
\begin{tabular}{@{}lcccc@{}}
\toprule
\multirow{2}{*}{\textbf{Fusion Strategy}}
& \multicolumn{2}{c}{\textbf{Beauty}}
& \multicolumn{2}{c}{\textbf{Clothing}} \\
\cmidrule(lr){2-3}\cmidrule(l){4-5}
& \textbf{N@20} & \textbf{H@20}
& \textbf{N@20} & \textbf{H@20} \\
\midrule
Last Layer                    & 0.0390 & 0.0989 & 0.0214 & 0.0542 \\
Mean Fusion                   & 0.0343 & 0.0842 & 0.0202 & 0.0531 \\
Random Fusion                 & 0.0332 & 0.0812 & 0.0197 & 0.0514 \\
\rowcolor[HTML]{96FFFB}
Increasing-depth ($\tau=0.8$) & \textbf{0.0394} & \textbf{0.0995}
                              & \textbf{0.0218} & \textbf{0.0578} \\
Increasing-depth ($\tau=1$)   & 0.0386 & 0.0974 & 0.0216 & 0.0552 \\
Descending-depth ($\tau=0.8$) & 0.0230 & 0.0549 & 0.0174 & 0.0432 \\
Descending-depth ($\tau=1$)   & 0.0286 & 0.0711 & 0.0183 & 0.0457 \\
\bottomrule
\end{tabular}
\end{table}

\subsection{Additional Item-Group Analysis}
\label{app:item-group-analysis}

To assess the item-level heterogeneity motivating Instance-Aware MoE, we analyze layer-wise patterns across semantic item groups.
This analysis is
based on the observation that embedding space geometry reflects semantic
relatedness among encoded texts~\cite{mikolov2013distributed,reimers2019sentence}.
Therefore, similar LLM representations may correspond to similar semantic
characteristics and layer-wise patterns.

We perform unsupervised clustering on LLM representations to obtain
coarse-grained semantic groups. We compare K-means, K-means++,
hierarchical clustering, and density-based clustering, and select
K-means++ considering embedding dimensionality, clustering stability, and
group size requirements.

To determine the number of item groups, we refer to the elbow method and the silhouette score. Based on these criteria, we set the number of clusters to six and apply K-means++ for clustering. For each item, we use its L2-normalized final-layer representation as the clustering input.
The resulting groups are used to compare how item representations evolve across LLM layers, providing the item-group evidence discussed in Motivation 3.

\subsection{Generalization Study}

To examine whether IMFuse transfers to a different LLM representation hierarchy, we use Qwen3-8B~\cite{qwen2025qwen3} to encode the same item text and use SASRec as the sequential recommendation backbone. By changing the LLM encoder, this experiment examines whether IMFuse can adapt its fusion strategy to a different set of intermediate-layer representations. As shown in \Cref{tab:qwen-generalization}, IMFuse improves all eight reported values, with average gains of 7.07\% for RLMRec and 4.46\% for SpecTran. These gains are consistent with the design of IMFuse: Adaptive Multi-Layer Fusion learns layer-aware and dimension-aware weights, while Instance-Aware MoE adjusts the contributions of shared experts for each item. The improvements across both semantic transformation pipelines suggest that IMFuse can make use of intermediate-layer information from Qwen3-8B as well as LLaMA-3-8B.

\begin{table}[t]
  \centering
  \caption{Generalization study with LLaMA-3-8B and Qwen3-8B.}
  \vspace{-0.3cm}
  \label{tab:qwen-generalization}
  \small
  \begin{tabular}{@{}cccccc@{}}
    \toprule
      & & \multicolumn{2}{c}{\textbf{Clothing}} & \multicolumn{2}{c}{\textbf{Toy}} \\
    \cmidrule(lr){3-4}\cmidrule(lr){5-6}
    \multirow{-2}{*}{\textbf{LLM Encoder}} & \multirow{-2}{*}{\textbf{Method}}
      & N@20 & H@20 & N@20 & H@20 \\
    \midrule
      & RLMRec & 0.0142 & 0.0357 & 0.0328 & 0.0811 \\
      & + IMFuse
        & 0.0148
        & 0.0396
        & 0.0350
        & 0.0863 \\
    \cmidrule(lr){2-6}
      & SpecTran & 0.0140 & 0.0468 & 0.0398 & 0.0977 \\
    \rowcolor[HTML]{96FFFB}
    \cellcolor{white}\multirow{-4}{*}{\textbf{Qwen3-8B}}
      & \cellcolor{white} + IMFuse
        & \textbf{0.0151}
        & \textbf{0.0489}
        & \textbf{0.0408}
        & \textbf{0.1006} \\
    \midrule
      & RLMRec & 0.0175 & 0.0435 & 0.0404 & 0.1006 \\
      & + IMFuse
        & 0.0179
        & 0.0453
        & 0.0413
        & 0.1035 \\
    \cmidrule(lr){2-6}
      & SpecTran & 0.0217 & 0.0572 & 0.0469 & 0.1189 \\
    \rowcolor[HTML]{96FFFB}
    \cellcolor{white}\multirow{-4}{*}{\textbf{LLaMA-3-8B}}
      & \cellcolor{white} + IMFuse
        & \textbf{0.0232}
        & \textbf{0.0605}
        & \textbf{0.0488}
        & \textbf{0.1219} \\
    \bottomrule
  \end{tabular}
  \vspace{-0.3cm}
\end{table}

\begin{figure}[t]
    \centering
    \includegraphics[width=\columnwidth]{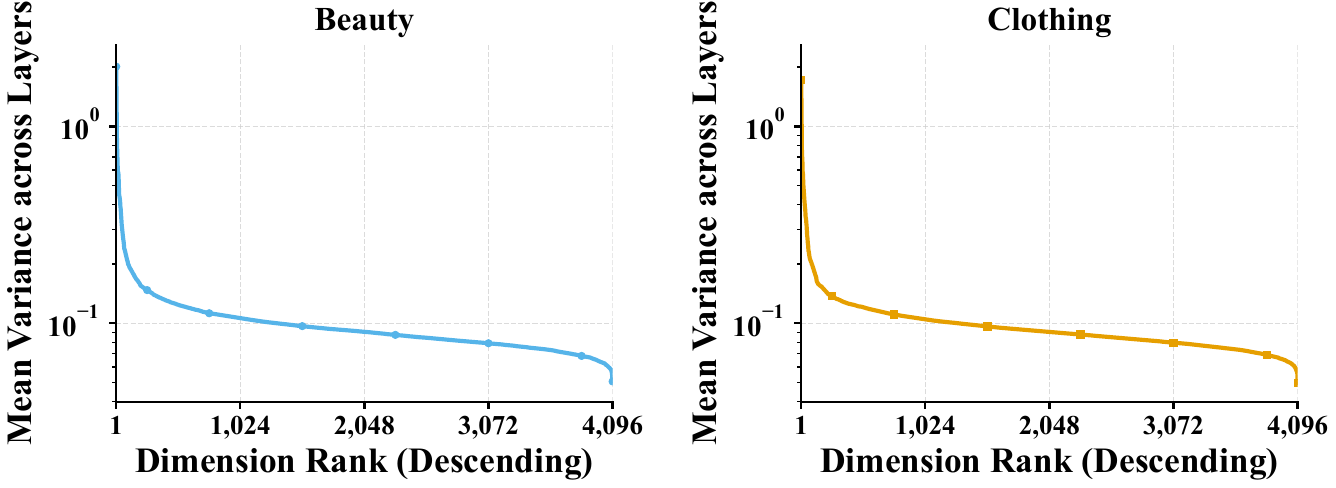}
    \caption{Dimension-wise variance of item embeddings on Beauty and Clothing, averaged over 33 LLM layers and sorted in descending order.}
    \label{fig:dimension-variance}
\end{figure}

\subsection{Dimension-wise Heterogeneity Analysis}
\label{app:dimension-analysis}

\textbf{Differences across layers.}
The results in Tables~\ref{tab:overall} and~\ref{tab:ablation} show that incorporating layer representations improves over using only the final layer. Moreover, changing the initialization of layer weights affects recommendation performance, indicating that how layer contributions are learned matters. Comparisons with alternative multi-layer fusion methods (Table~\ref{tab:fusion-comparison}) further highlight the effect of different aggregation strategies. Together, these results suggest that LLM layers provide useful but unequal contributions to recommendation and that IMFuse learns more effective fusion weights for combining them.

\textbf{Differences within layers.}
Beyond layer-level differences, dimensions within an item embedding can exhibit different statistical properties. LLMInit \cite{zhang2025llminit} uses variance across items as a heuristic for identifying informative dimensions and selects the highest-variance dimensions for recommendation initialization. As shown in Figure~\ref{fig:dimension-variance}, our analysis reveals substantial differences in variance across dimensions, even after averaging over all 33 LLM layers. This observation motivates treating embedding dimensions differently rather than assuming identical contributions, providing a rationale for our dimension-wise fusion weights.

\end{document}